\documentclass[fleqn,usenatbib]{mnras}

\usepackage{newtxtext,newtxmath}

\usepackage[T1]{fontenc}

\DeclareRobustCommand{\VAN}[3]{#2}
\let\VANthebibliography\thebibliography
\def\thebibliography{\DeclareRobustCommand{\VAN}[3]{##3}\VANthebibliography}

\usepackage[mathscr]{euscript}
\usepackage{graphicx}	
\usepackage{amsmath}
\usepackage{bm}
\usepackage[dvipsnames]{xcolor}
\usepackage[nice]{nicefrac}
\newcommand{\orcid}[1]{\href{https://orcid.org/#1}{\includegraphics[width=8pt]{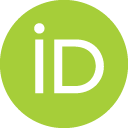}}}
\usepackage{subcaption}
\usepackage{caption}
\usepackage{tabularx}
\usepackage{booktabs}
\usepackage{amsfonts}
\usepackage{mathrsfs}

\title[planetesimals in Binaries]{Capture of Interstellar Planetesimals in Protoplanetary Discs in Binary Star Systems}

\author[Pollard et al.]{
Alia Pollard$^{1}$\thanks{E-mail: alia.pollard1@monash.edu}
\orcid{0009-0006-7548-833X}
Evgeni Grishin$^{1,2}$\thanks{E-mail: evgeni.grishin@monash.edu.au} 
\orcid{0000-0001-7113-723X}
\\
$^{1}$School of Physics and Astronomy, Monash University, Clayton, VIC 3800, Australia\\
$^{2}$OzGrav: Australian Research Council Centre of Excellence for Gravitational Wave Discovery, Clayton, VIC 3800, Australia
}

\date{Accepted XXX. Received YYY; in original form ZZZ}

\pubyear{2025}

\begin{document}
\label{firstpage}
\pagerange{\pageref{firstpage}--\pageref{lastpage}}
\maketitle

\begin{abstract}
Planet formation requires growth over 12 orders of magnitude in size, from micron-sized dust grains to fully formed planets. The formation of $\sim \rm km$ sized planetesimals is the most challenging stage, where radial drift and fragmentation rapidly remove meter-sized solids from the protoplanetary disc. Overcoming this barrier requires favourable conditions for streaming instability to concentrate solids. These problems are exacerbated in binary star systems, where gravitational perturbations and disc truncation reduce the efficiency of planetesimal formation pathways. Planetesimal capture provides an alternative mechanism where unbound planetesimals are captured by the protoplanetary disc through gas-drag, potentially bypassing the most restrictive growth stages. We investigate this process in circumstellar and circumbinary discs using over $3\times10^7$ numerical experiments across a range of binary separations. For circumstellar discs, capture of large ($\gtrsim 50$ m) planetesimals is suppressed below $\sim 50$ au, with the strongest reduction at $\sim 10$ au, broadly consistent with the observed decrease in planet occurrence in binaries. Tight ($\sim 10\ \rm au$) binaries can still capture $\mathcal{O}(1)$ km-sized planetesimals, increasing to $\mathcal{O}(10)$ km-sized in wider systems during the birth cluster's lifetime. Circumbinary discs are much less efficient, with capture rates dropping to $\sim 1\%$ of the single-star case for binaries above $\sim 1$ au and becoming negligible for large planetesimals at wider separations. Overall, planetesimal capture provides a supplementary channel for delivering solids to binary systems and mitigating inefficiencies in in-situ planetesimal formation.
\end{abstract}

\begin{keywords}
planets and satellites: formation -- comets: general -- minor planets, asteroids:
general -- minor planets, asteroids: individual: 1I/2017 U1 (’Oumuamua)
\end{keywords}

\section{Introduction} \label{sec:intro}

Despite the abundance of exoplanetary systems, the theory of planet formation is far from understood. Planet formation requires growth by $12$ orders of magnitude in size from micron-sized dust grains to fully formed giant planets 
 within less than few Myr \citep[e.g.][]{raymond22}. While the growth of grains ($\sim1\mu$m) to boulders ($\sim1-10 $m) and the growth of $>\rm km$-sized planetesimals ($\sim1-100$km) to planets is relatively robust via sticking and pebble accretion, respectively \citep{lambrechts12, blum18, armitage_book}, formation of the first planetesimals is challenging. In particular, pebbles ($\sim 1$mm) that grow to cm-m sizes are decoupled from the gas and experience a radial drift which causes a rapid in-spiral, known as the radial drift (or meter size) barrier \citep{wei77}. The maximal radial drift speed can reach up to $10^4\ \rm cm/s \sim 10^{-2}\ \rm au\ yr^{-1}$, leading to an in-spiral times of $\sim 10^3\ \rm yr$ at $10$ au. The loss of the pebbles occurs faster than planetesimals form, which is detrimental to planet formation.

Several pathways of overcoming the meter barrier have been suggested, including successive growth due to repeated collisions \citep{windmark12} and local over-density and direct gravitational collapse due to the streaming instability \citep{youdin2005, 2007Natur.448.1022J}. Streaming instability has been successful in the outer zones of our solar system near the Kuiper-belt \citep{kbo19}. However, it also suffers from shortcomings: i) Early simulations have shown successive gravitational collapse for same size particles. However, for a mass spectrum, collapse was not achieved and the simulation did not converge \cite{krapp19}. ii) `Optimal' pebble sizes and metal-rich discs are required to achieve the conditions for gravitational collapse (i.e. local dust to gas ratio over unity, \citealp{li-youdin21}). Many recent detailed simulations include 3D, non-linear coupling and mass spectrum \citep[e.g.,][]{si-code-comparison}. While 3D simulations attain strong clumping at lower metallicities, the fate of the clumped particles is less clear, and further analytical and numerical studies are required to estimate the efficiency of the streaming instability in the diverse setting of planet formation. 

The above suggests that regardless of the exact conditions required for the streaming instability to operate, additional processes are required to work in tandem with the streaming instability to successfully form planets at the observed occurrence rates.
Planetesimal capture, a mechanism proposed by \cite{gri19}, allows for larger interstellar (or intra-cluster) planetesimals to be captured in protoplanetary discs. Planetesimal capture can also occur in other environments such as molecular clouds and star forming cloud cores \citep{seeding1, seeding2, seeding3}.

While planetesimals capture cannot explain the formation of the first planetesimals (which could form via streaming instability as a rare, fine-tuned event), it allows effective \textit{planetesimal seeding} which can lead to successful planet formation in the young protoplanetary discs of the young stellar birth cluster. Aeolian erosion \citep{erosion1, erosion2} can additionally sort pebbles by their size, assisting both streaming instability and pebble accretion. However, erosion may not affect the dust evolution when radial drift, fragmentation and dust traps are present \citep{erosion3}.

Even with the challenges inherent to planet formation, more than $6,000$ exoplanets have been discovered, exhibiting a wide range of orbital architectures, with many residing in multi-planetary systems \citep[e.g., ][]{cloutier26}. The fraction of planets in binary stellar systems has been systematically investigated. Earlier estimates suggested a relatively low occurrence rate of $\sim 20\%$ \citep{fonta21, gonzalez24}, although radial-velocity (RV) samples are affected by significant observational biases. More recently, \cite{thebault2025} presented a more complete census of planets in binary systems, finding a $24\%-34\%$ planet multiplicity fraction around GFA primaries within $500\ \rm pc$ (and a total multiplicity fraction of $22.5\%$). Although this remains lower than the field star multiplicity of $33\pm2\%$ \citep{raghavan2010}, planets in binary systems are nevertheless common.

Building on this, \cite{maxwell2021} compiled results from multiple radial-velocity and imaging surveys and identified a clear deficit of planets in binaries with separations $\le 200^{\ +200}_{-120}$ au. In particular, only $\sim 15^{\ +17}_{-12}\%$ of binaries with separations around $\sim 10$ au host planets compared to single stars. Consistently, the separation distribution of planet-hosting binaries peaks at a few hundreds of au, whereas the field binary population peaks at much smaller separations of $30$–$40$ au \citep{raghavan2010}.

The formation of these planets in binary systems is compounded by challenges not present in single star systems. Circumstellar discs in binary systems can be truncated by the effect of the companion, and circumbinary discs have cavities at their centre \citep{artymowicz1994}. Consequently, discs in binary systems have less material than discs in single systems, hindering the conditions for successful planet formation. Binaries also contain regions of instability where planet orbits cannot be maintained, leading to gaps with no possibility for planets to form \citep{holman1999}. Because of these factors, we expect planets to be rarer in binaries than in single systems \citep{thebault2025}. This motivates further research into planet formation in binary systems, as exposing exactly how planets can form in such restrictive environments helps illuminate the requirements for planet formation generally.

Discovering viable mechanisms for binary planet formation is crucial to an overall understanding of planet formation. Planetesimal capture has been previously tested for single systems by \cite{gri19}, but remains untested for binaries. This paper aims to test the possibility of planetesimal capture in binary systems. We use an equal mass binary of sun-like stars and vary the separations to explore the relationship between binary separation and capture rates. We test capture for circumstellar discs around each star and a single circumbinary disc around both stars. 

The paper is organised as follows: in Section \ref{sec:2} we compare binary capture statistics to the capture statistics around single stars using a simplified comparative model (Sections \ref{sec:2.2} and \ref{sec:2.3}). In Section \ref{sec:3} we present the results of our main model for systems with circumstellar discs around both stars (Section \ref{sec:3.1}) and circumbinary discs (Section \ref{sec:3.2}). In Section \ref{sec:4} we discuss the model and its implications, and summarise our findings in Section \ref{sec:5}. 

\section{Comparison to capture around single stars}\label{sec:2}
Here we overview the main results of the semi-analytical model of \cite{gri19} for capture around single stars. First, we overview the theory behind single star capture. We then compare single star capture results to a model created specifically for the purpose of comparison.
\subsection{Overview of capture around single stars}\label{21}

The protoplanetary disc consists of mostly gas and only about $1\%$ of dust \citep{rosetta}. The main interaction of a planetesimal passing through the disc is aerodynamic gas drag that may lead to the capture of the planetesimal. We assume the gravity of the disc is negligible compared to the central star, such that the disc is gravitationally stable.
The force of the drag force is \citep{wei77}
\begin{equation}\label{Dragforce}
    \boldsymbol{F}_\text{\rm D}=-\frac12 C_\text{D}\pi R_\text{p}^2 \rho_\text{g}v_\text{rel}^2 \bm{{{\hat{v}_{\text{rel}}}}}\text{,}
\end{equation}
where $C_\text{\rm D}$ is the dimensionless drag coefficient, $R_\text{p}$ is the radius of the planetesimal, $\rho_\text{g}$ is the density of the disc, and $v_{\text{rel}}$ and its unit vector counterpart describe the magnitude and direction of the relative velocity of the planetesimal within the disc. The characteristic stopping time for a particle is $t_{\rm s}=mv_{\rm rel}\big/|\boldsymbol{F}_{\rm D}|$. Small grains have short stopping times compared to the orbital period, $t_{\rm s} \ll P$, while larger planetesimals have much longer stopping times $t_{\rm s} \gg P$. The pebbles with $t_{\rm s} \sim P$ are the ones that experience the fastest radial drift.

During one perpendicular passage through the disc the drag force applied on the planetesimal changes the orbital energy by
\begin{align}
\Delta E & =  \int \boldsymbol{F}_{\rm D} \cdot \bm{v}_\text{rel} \text{ d}t \nonumber \\
 & = -\frac{C_\text{D}\pi R_\text{p}^2\rho_0}{2}\int^{\infty}_{-\infty}\exp\left(-\frac{v_{\text{rel}}^2t^2}{2h^2}\right)v_\text{rel}^3 \text{ d}t \nonumber  \\
 & \approx  -\frac{C_\text{D}\pi R_\text{p}^2\Sigma_\text{g}v_\text{rel}^2}{2}\text{,}     \label{work}
\end{align}
where $\bar a$ is the radial distance and $\rho_0=\rho_g(\bar{a}, z=0)$ is the mid-plane density at $\bar{a}$ \citep{gri19}. 
The trajectory is integrated over the time the planetesimal spends in the disc. $h$ is  the scale height of the disc and we assume the relative velocity is constant during the passage (see \citealp{gri19} for details). $\Sigma_\text{g}$ is the radial gas surface density of the disc,  
\begin{equation}\label{surfacedensity}
    \Sigma_\text{g}=\Sigma_\text{g, 0}\left( \frac{a}{\text{au}} \right)^{-\beta}\text{,}
\end{equation}
where $\Sigma_\text{g, 0}$ is the central radial gas surface density, and $\beta$ is a scaling factor for the dependence of radial surface density on position in the disc.

We follow the power law from the canonical minimum mass solar nebula scaling is $\beta=3/2$ \citep{h81,mmsn}, and choose a flat aspect ratio $h/a=10^{-2}$ for simplicity. The volume density is 


\begin{equation}\label{rhog}
    \rho_{\text{g}}=\frac{\Sigma_g}{\sqrt{2\pi h}}\rho_0\left(\frac{a}{\text{au}}\right)^{-5/2}\exp \left( {\frac{-z^2}{2h^2}} \right)\text{,}
\end{equation}
where the normalisation is due to integrating the density over a Gaussian vertical profile, $\Sigma_g = \int_{-\infty}^{\infty}\rho dz$, which is a product of hydrostatic equilibrium \citep[e.g.,][]{armitage2015}. We set $\rho_0 = 10^{-8}\ \rm g\ cm^{-3}$, which is a factor of a few larger that he MMSN value, consistent with the diversity of Kepler planets and the average observed value of the solid component of the surface density \citep[see e.g.,][for details]{raymond14}.
The scale height $h$ is related to the sound speed $c_s$ and the Keplerian frequency $\Omega$  via $h=c_s/\Omega$. The drag coefficient is adaptive throughout the disc model and changes according to the Reynolds number $\texttt{Re}$. It is given by  

\begin{equation}\label{DragCoeff}
    C_\text{D}=\frac{24}{\texttt{Re}}(1+0.27\texttt{Re})^{0.43}+0.47\left[1-\exp \left( {-0.04\texttt{Re}^{0.38}} \right) \right] \text{,}
\end{equation}
a prescription taken from \cite{perets2011}. The Reynolds number is given by:
\begin{equation}\label{Re}
    \texttt{Re}=\frac{4R_{\text{p}}v_{\text{rel}}}{\bar{v}_{\text{th}}\lambda}\text{.}
\end{equation}
The relative velocity $v_{\rm rel}$ depends on the orbit, the mean free path ($\lambda$) is  $\lambda=({n_g\sigma})^{-1}$, and the mean thermal velocity is $\bar{v}_{\text{th}}=\sqrt{8/\pi}c_s$.
The gas number density is $n_g=\rho_g/\mu$, the neutral gas-gas collision cross section is $\sigma=10^{-15}\ \text{cm}^2$, and  $\mu=2.3m_H$ is the mean molecular weight \citep{perets2011}. 
Together, this adaptive drag coefficient along with the density profile create a model of the disc that can be applied to N-body simulations where it acts as an external force.

Planetesimals entering this simulated disc will experience the drag force and may loose enough energy to be captured. For an initial velocity at infinity $v_\infty$, the kinetic energy is $m_pv_\infty^2/2$. Capture will occur if the energy loss exceeds the initial kinetic energy, namely $\Delta E > m_p v_{\infty}^2/2$. A version of this condition for planetesimal capture in a single star system was developed by \cite{gri19}, and is given by 
\begin{equation}\label{capcondition}
    R_\text{p}\lesssim\frac34\frac{C_\text{D}\Sigma_\text{g}}{\rho_\text{p}}\left(1+\frac{v_{\rm esc}^2}{v_{\rm \infty}^2}\right) \equiv \frac34\frac{C_\text{D}\Sigma_\text{g}}{\rho_\text{p}}(1+\Theta_\text{s})
\end{equation}
where $\rho_\text{p}$ is the planetesimal density, and $\Theta_\text{s} \equiv (v_{\rm esc}/v_{\rm \infty})^2$ is the Safronov number.

Based on this condition, the possibility of capture depends on the size of the planetesimal, its velocity at infinity, and its escape velocity. Planetesimals that interact closer to the star have higher escape velocities, as they experience a stronger gravitational force from the star and more drag from the dense inner disc. 

For the gravitationally focused regime, $v_\infty\ll v_\text{esc}$ ($\Theta_s \gg 1$), the trajectory is significantly deflected by the star, leading to passage at the inner, denser zones in the disc. This regime dominates for larger planetesimals that have inherently more energy when they enter the system, and therefore require closer encounters with the star to be captured. For the geometric regime, $v_\infty\gg v_\text{esc}$ ($\Theta_s \ll 1$), the gravity of the star is negligible and the orbit is almost unperturbed. This regime is dominant for smaller planetesimals that can be easily captured in the edge of the disc.

For a critical size, a transition occurs between both regimes where the capture probabilities are roughly equal.
This should not be affected by the presence of a binary star, as is evident in the results (sec. \ref{sec:3}). 

Capture in binary stars poses additional challenges. For the perturbative regime, where the binary companion is much further than the radial location of the impending planetesimal, we expect to see similar results to the single case. However, in the general case where the planetesimal's orbit is significantly altered by a closer companion star, the trajectories become non-integrable, and the centre of mass is different from the centre of each disc, which can also be truncated compared to the single case. Thus, we will estimate the efficiency of capture in binary systems numerically, as described below. To determine how different binary systems are to single systems, we construct a comparative model with the aim of directly comparing single and binary capture under similar initial conditions.


\subsection{Setting up a simplified comparative model} 
\label{sec:2.2}

\begin{figure}
    \centering
    \includegraphics[width=0.9\columnwidth]{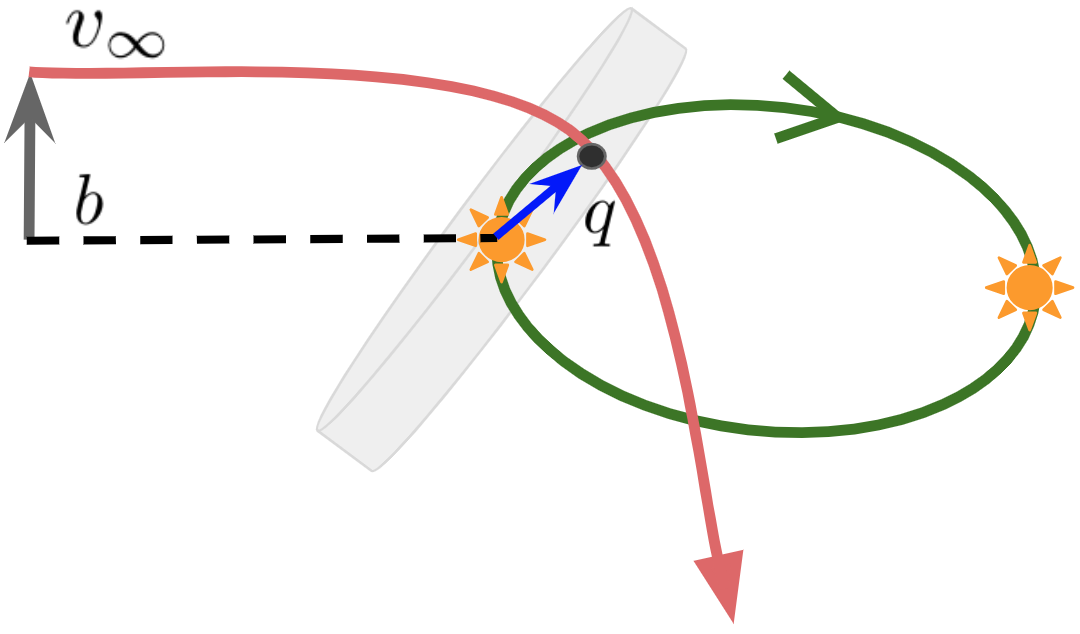}
    \caption{The set up for the comparative model. A planetesimal is assigned an impact parameter (b) and velocity at infinity ($v_\infty$) and aimed at the star in the binary with a disc at peri-centre $q$.}
    \label{fig:Toymodelschematic}
\end{figure}

\begin{figure*}
    \centering   \includegraphics[width=0.83\linewidth]{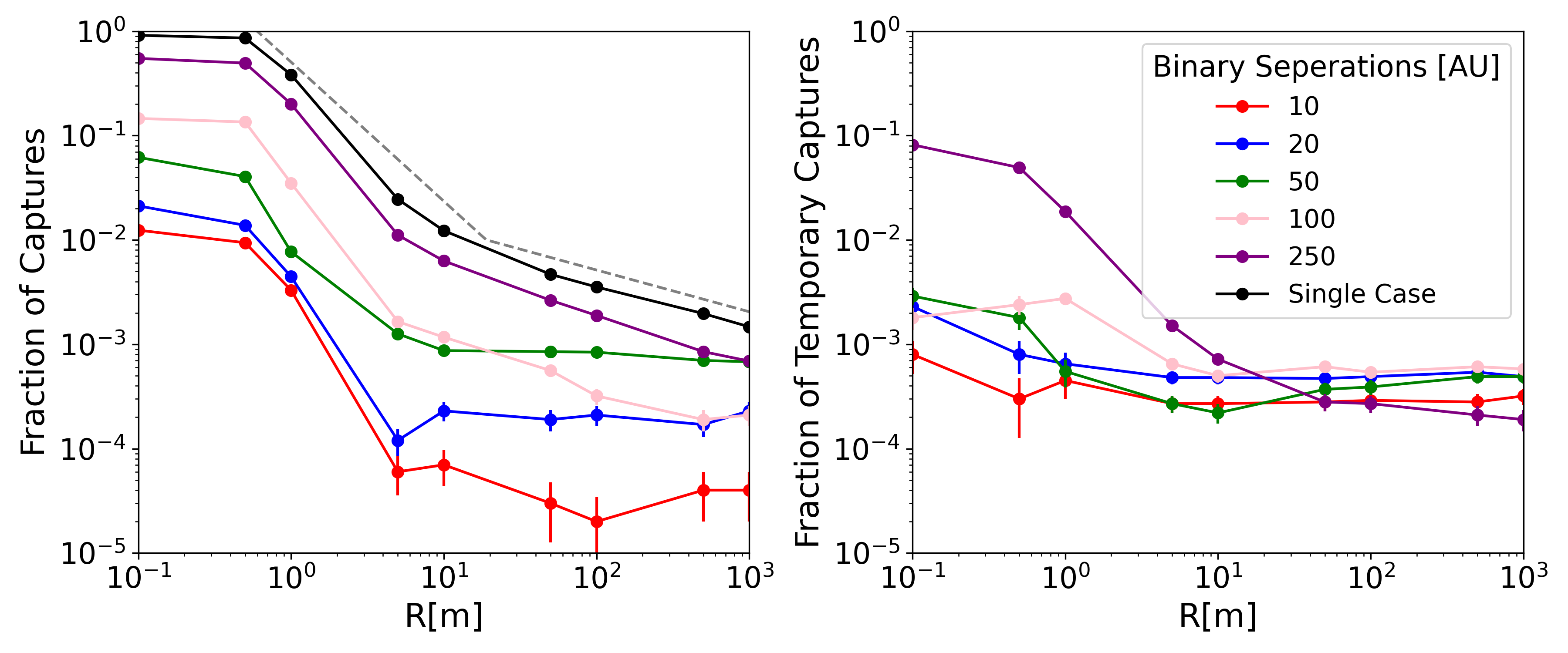}
    \caption{Fraction of planetesimal interactions that resulted in a capture (left) or a temporary capture (right) for planetesimals of different sizes (R) and stars with different separations. Each point represents anywhere from 10,000-100,000 simulations. The single case results were also obtained with REBOUND using the same set-up for direct comparison. The grey dotted line represents the theoretical amount of captures for the single case derived by \citet{gri19}.}
    \label{fig:toymodelfractionresults}
\end{figure*}
\begin{figure}
    \centering
 \includegraphics[width=0.97\linewidth]{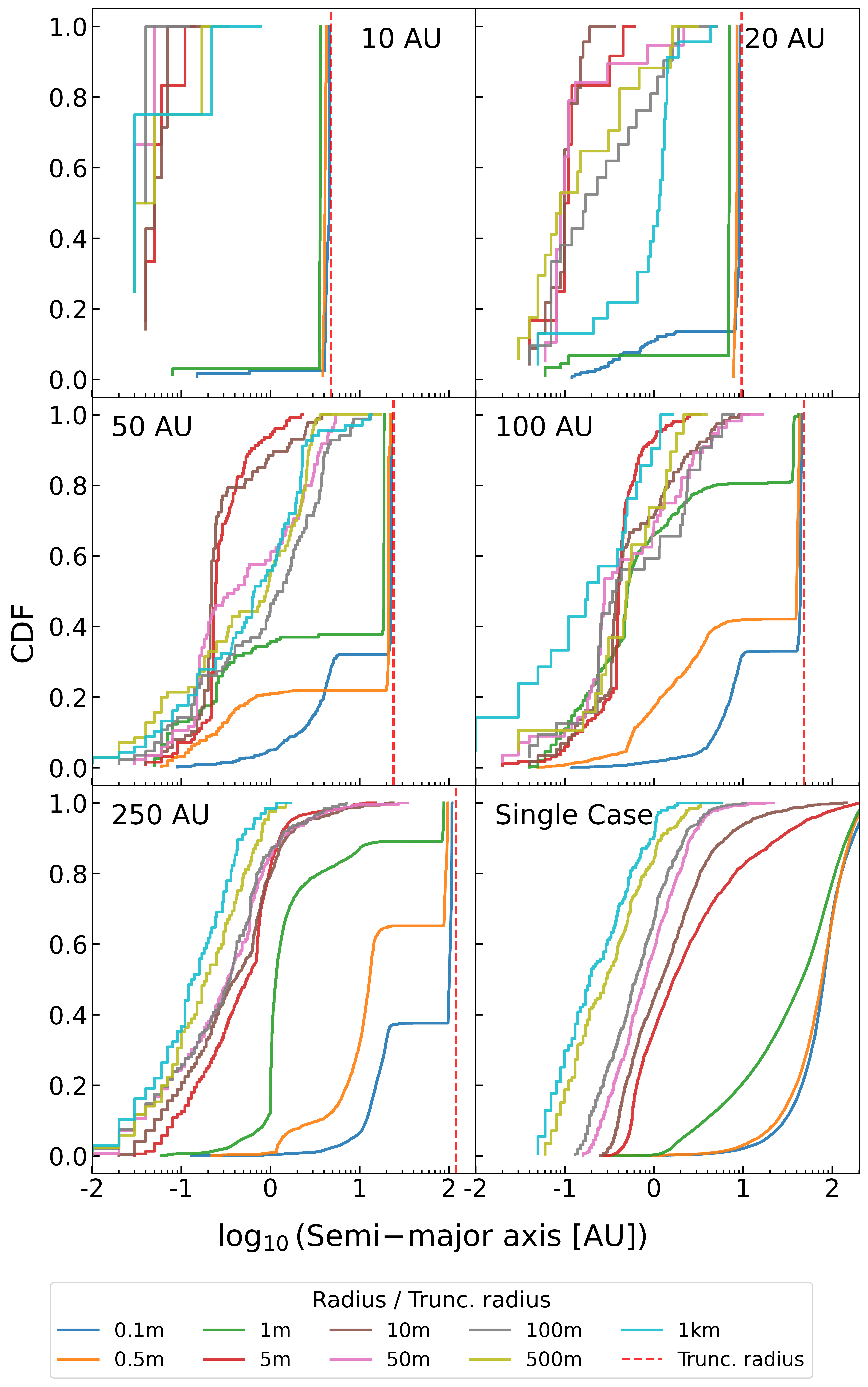}
    \caption{Cumulative distribution functions for the final semi-major axis values of captured planetesimals for the comparative model. Each panel shows a different binary separation 
    with lines for each planetesimal size. The red dotted line indicates the truncation radius of the disc for that separation. The final panel is the single case for comparison.}
    \label{fig:CDFToy}
\end{figure}

We initially created a comparative model to more directly compare to the single star results obtained by \cite{gri19}. Our comparative model is of a system with only one circumstellar disc and a binary companion, where the planetesimal is always aimed at the star with the disc. This set up is displayed in Figure \ref{fig:Toymodelschematic}.

We treat the binary star as a perturber, allowing for direct single case comparison. The planetesimal starts at peri-centre $q$ and evolves backward in time to a sufficiently large distance of 1000 au with the drag force turned off to avoid skewing the position of the planetesimal. After that we turn on the gas drag from the disc and evolve the system forward in time.

We use the \texttt{IAS15} integrator \citep{ias15} to run N-body simulations with two stars, one of which has a disc surrounding it. Our disc is initialised with an outer radius of $250$ au, which is then truncated by the presence of the binary star in accordance with \cite{artymowicz1994}
\begin{equation}\label{truncradius}
    r=0.48a 
\end{equation}
for equal mass, circular binaries which we use in our simulations. Varying the mass ratio and eccentricity is beyond the scope of this paper and will be studied in future work.
To quantify the effects of disc truncation, we simulate different models for stellar separations at $10, 20, 50, 100,$ and $250$ au, and for planetesimals of nine different radii ranging between $10\ \rm cm$ and $1\ \rm km$. These planetesimals all have a density of $\rho_\text{p}=1\text{ g/cm}^3$. Each planetesimal size was run for each radius, with a minimum of 10,000 runs, where larger planetesimals had a larger number of runs to increase the number of captured planetesimals and improve the statistics (see \ref{appen:a} for details).

The planetesimals were all initialised with Monte-Carlo sampled orbital elements. The orbital angles are uniformly sampled. The cosine of the inclination was also uniformly sampled, and from it the inclination was determined. The impact parameter is sampled from an area-uniform distribution, biasing higher impact parameters with a maximum of $b_{\rm max} = 100\ \rm au$. The velocities are taken from a $\chi(4)$ distribution using a velocity dispersion value of $6.2\ \rm km/s$. This value was chosen to match typical object ejection velocities from planetary systems \citep{Adams05}. 

In the binary case, other outcomes beyond the single case's capture/no capture are possible. Three-body dynamical interactions can occur, causing a temporary capture as opposed to the drag force induced captures. These temporary captures are usually resolved within several orbital periods and remain highly eccentric throughout the evolution. To distinguish between the two, orbits in this model were run for longer after their eccentricity dips below one, and if after $30$ orbits their eccentricity was below $0.1$ they were classified as true captures. We've chosen these thresholds to balance between the need to distinguish between true and temporary capture and to remain computationally efficient by avoiding simulation long-term expensive circularisation and migration. These simulations continued until their eccentricity decreased below this level and they were captured, or they were ejected and classified as a temporary capture.

\subsection{Results of the comparative model} \label{sec:2.3}
The fraction of simulated planetesimals that were captured or temporarily captured is shown in Figure \ref{fig:toymodelfractionresults}. The results of the simulated single star case are also plotted for direct comparison with the binary captures. The presence of a binary is detrimental to the possibility of capture. At the $250$ au separation, the results are similar to the single case, but with less captures. Theoretically, each closer separation should further decrease capture, which is the general trend shown from the comparative model, with an exception in the 100 au separation. This is a numerical artifact of choosing $b_{\rm max}$ that matches the binary separation of $100$ au. This choice causes many of the planetesimals to have gravitational interactions with the binary star, decreasing the capture rates. Gravity of the binary is an expected barrier to capture, although the specific coupling here of the planetesimals typical trajectory coinciding with the companions orbit artificially amplifies that effect in simulations as opposed to a true environment. 

The right panel of Figure \ref{fig:toymodelfractionresults} shows the amount of temporary captured is relatively flat and independent of the planetesimal size (for planetesimals above $10\ \rm m$) and binary distance. This trend is best explained by the temporary captures being orbits that have lost just enough energy to be bound, but settled in the outskirts of the disc where the gas density is low and binary perturbation is strongest. Hence, they cannot circularise effectively, but rather are perturbed to ejection by the binary companion.

Figure \ref{fig:CDFToy} displays cumulative distributions for the final semi-major axis values of captured planetesimals. The flat regions on the CDF lines indicate unstable regions, and are consistent with the inner stability limit circumstellar planet of $a_c\approx 0.27a_{\rm bin}$ for equal mass circular binaries \citep{holman1999}. Planetesimals that become bound due to gas drag at these locations are unable to retain dynamically stable, and no captures are recorded ending at those a values. Some small pebbles  ($\le1\ \rm m$) are able to retain an elongated orbit near the disc edge where they are recaptured efficiently to a circular orbit, only to be ejected by the companion when they are closest to the centre of mass. This can occur only if the stopping time $t_{\rm s}$ is shorter than the orbital period. Larger planetesimals are effectively ejected since they cannot circularise rapidly enough. No planetesimals are captured beyond the truncation radius, as there is no disc there to capture them. 

The single case provides valuable comparison to the binary case. The flat unstable regions are a product of the binary and, as expected, they do not exist in the single case. It also can be seen that generally larger planetesimals settle at lower semi-major axis values, because they often require a closer encounter with the star and the denser regions of the disc to be captured. The comparative model is a good tool for comparison to the single case. We also explore a model in sec. \ref{sec:3} where both stars have their own discs (in the circumstellar case) and the orbit at infinity is assigned with the respect to the centre of mass of the binary.
\section{Population study}\label{sec:3}

\begin{figure}
    \centering
    \includegraphics[width=0.9\columnwidth]{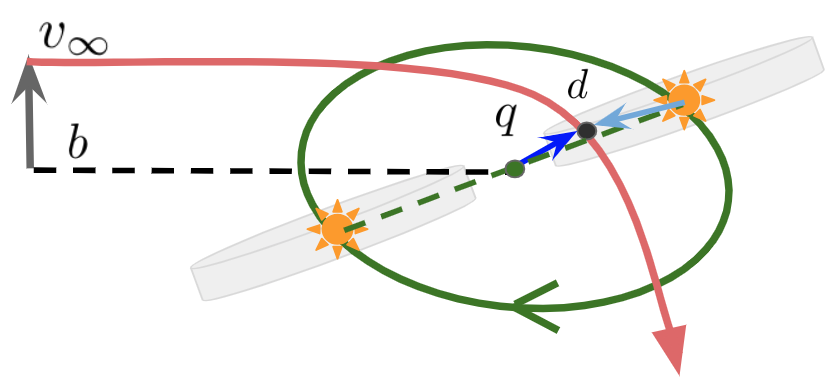}
    \caption{Sketch of the physical model. The orbit is initialised with velocity $v_\infty$ and impact parameter $b$ with respect to the centre of mass of the binary star. Both stars have discs. The peri-centre $q$ is measured from the centre of mass, which is different from the radial distance within the disc $d$ that the planetesimal is passing.}
    \label{fig:physical model}
\end{figure}

\begin{figure*}
    \centering
    \includegraphics[width=0.84\textwidth]{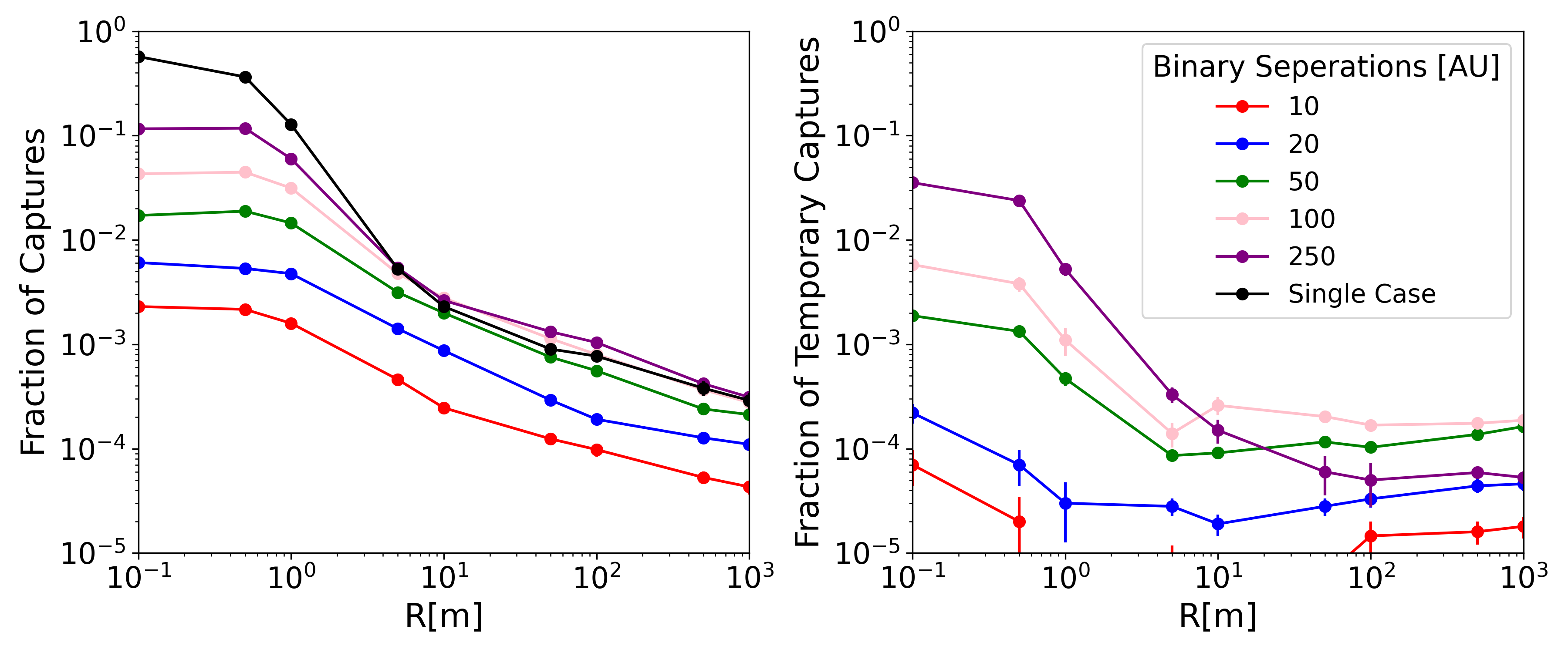}
    \caption{The same as Figure \ref{fig:toymodelfractionresults}, but for the physical model. The single case was rerun for $b_{\rm max}=250$ and plotted for visual comparison, although it is not directly comparable as can be seen when binary captures surpass the single case at some radius values.}
    \label{fig:physicalcapturefraction}
\end{figure*}
\begin{figure}
    \centering
    \includegraphics[width=0.98\linewidth]{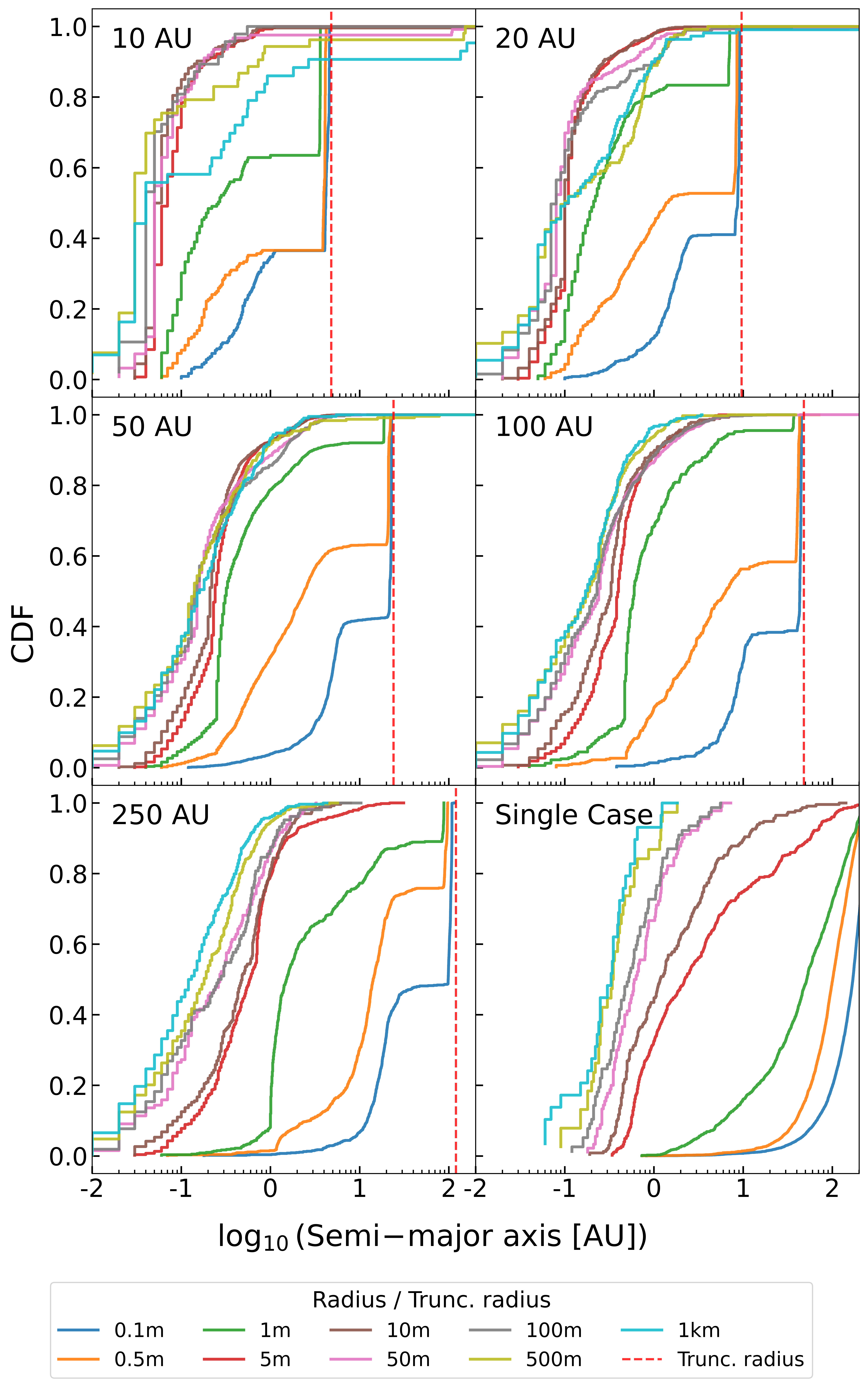}
    \caption{The same as Figure \ref{fig:CDFToy}, but for the physical model. The single case plotted here is for $b_{\rm max}=250$ to match the value used for the binary panels.}
    \label{fig:CDFphysical}
\end{figure}
In this section, we explore the main, physically motivated model. Here we describe the set-up of the circumstellar case, where the circumbinary case is described in sec. \ref{sec:3.2}. 
\begin{figure*}
    \centering
    \includegraphics[width=0.84\linewidth]{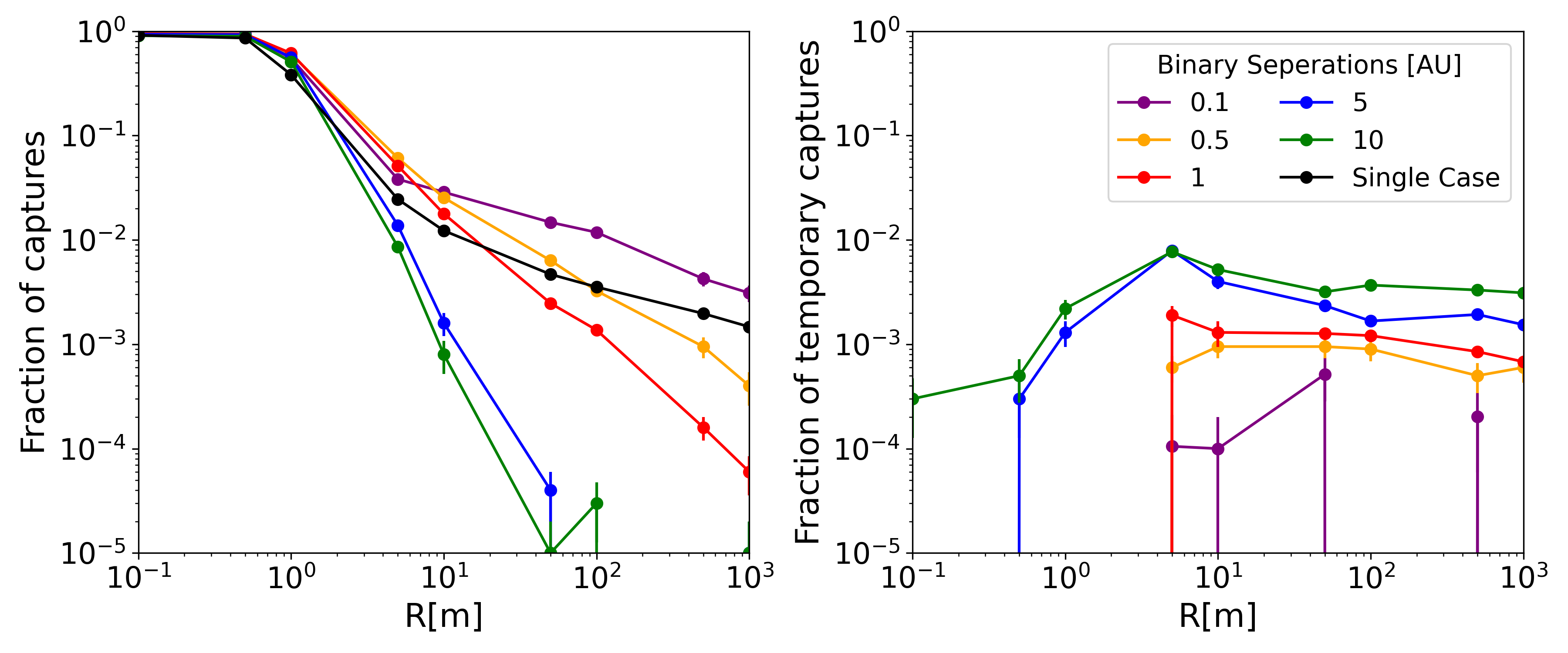}
    \caption{Capture and temporary capture fractions for Circumbinary systems of separations 0.1, 0.5, 1, 5, and 10 au compared to the single star case. Large planetesimals struggle to be captured at larger circumbinary separations due to the lack of an inner disc.}
    \label{fig:CBfracs}
\end{figure*}
\begin{figure}
    \centering
    \includegraphics[width=0.98\linewidth]{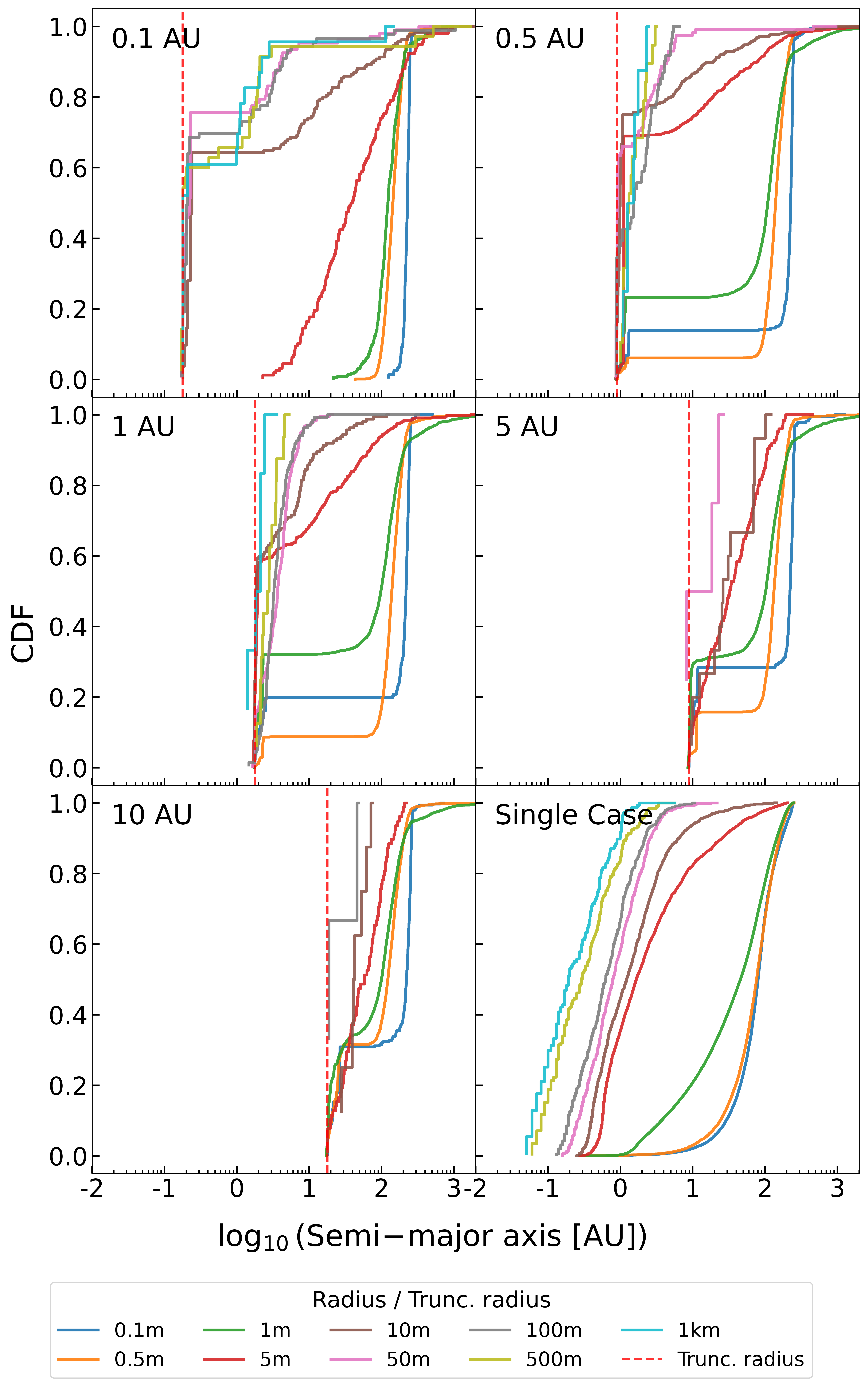}
    \caption{Cumulative distribution function for the final a values of captured planetesimals in circumbinary systems. As in the circumstellar case, unstable regions in the disc are visible in the flat sections.}
    \label{fig:CBCDF}
\end{figure}

\subsection{Circumstellar case}\label{sec:3.1}

For the circumstellar case, we model a circular equal-mass binary with two $1\ M_\odot$ stars, each surrounded by a circumstellar disc truncated according to the same prescription used in the comparative model. The maximal impact parameter is set to $b_{\rm max} = 250\ \rm au$, measured from the centre of mass (COM) of the binary. This ensures that the full geometry is captured even for the widest binary separation of $250$ au; otherwise, many planetesimals would pass between the stars for smaller impact parameters. Increasing $b_{\rm max}$ reduces the capture rate, since it scales as $b_{\rm max}^{-2}$, and therefore we ran more simulations to obtain good statistics. The distributions of impact parameter $b$, velocity at infinity $v_\infty$, and orbital angles are sampled identically to the comparative model.

Unlike the single-star case, planetesimals are aimed at the COM of the binary rather than at an individual star. Consequently, the impact parameter refers to the closest approach to the COM. This geometry is illustrated in Figure \ref{fig:physical model}. Because of this difference, the binary and single-star models are not directly equivalent, although the single-star case is still shown in Figures \ref{fig:physicalcapturefraction} and \ref{fig:CDFphysical} for comparison.

Some binary configurations exhibit higher total capture fractions than the single-star case as seen in Figure \ref{fig:physicalcapturefraction}. This does not imply that binaries generically capture more planetesimals; rather, the presence of two circumstellar discs provides more locations where planetesimals can encounter dense gas and dissipate energy. When considering captures onto only one binary component as in Figure \ref{fig:toymodelfractionresults}, the single-star case consistently produces higher capture fractions than the binaries.

The circumstellar-disc model produces capture fractions that follow trends similar to the single-star case even at small binary separations. Tighter binaries still exhibit lower capture efficiencies due to stronger disc truncation, although even binaries with separations as small as $10$ au retain a non-zero probability of capture. At larger planetesimal radii, the capture fractions become more similar across different binary separations because gravitational focusing (see sec. \ref{21}) dominates the dynamics. In this regime, planetesimals must pass close to a star to be captured, making the outer disc structure less important. Consequently, disc truncation has a weaker effect on large planetesimals, since capture mainly occurs within the dense inner disc regions.

Figure \ref{fig:CDFphysical} shows the final semi-major axis values of captured planetesimals for the physical model. Similarly to the comparative model, unstable regions of the discs are visible in the flat portions of the CDF. For smaller separations, these regions are more prominent than in the comparative model due to the increased captures. Smaller planetesimals that are being pushed outwards from the unstable regions are escaping the systems and becoming temporary captures rather than settling, or are exhibiting similar behaviour to the outer captures in the comparative model as described in Section \ref{sec:2.3}. This matches with the increase in temporary captures seen in Figure \ref{fig:physicalcapturefraction} at the smaller planetesimal radius values.

This model shows that the general conclusions drawn from the comparative model are robust, and additionally provides more clarity into capture in binaries. For circumstellar discs, it shows that capture is primarily affected by the binary separation and (as previously shown in \citealp{gri19}) planetesimal size. It also further highlights the effect of unstable regions in circumstellar discs on planetesimal capture as seen in Figure \ref{fig:CDFphysical}.

\subsection{Circumbinary case}\label{sec:3.2}

Circumbinary planets are much less common, with only 31 currently known \citep{thebault2025}. Most circumbinary planets tend to orbit tight binaries of separations around $0.1$ au, while the widest circumbinary planet orbits a $\sim 8$ au binary \citep{thebault2025}. 
Circumbinary discs have an inner truncation radius where the binary carves out a cavity. We used the conservative estimate of \cite{artymowicz1994}, which places the inner edge of the disc at $1.78$ times the binary separation. This inner truncation makes capture of larger planetesimals much harder at larger separations, as they typically need to enter inner regions of the disc to be captured. 

In order to explore the binary separation space, we run our code for the separation values of $0.1, 0.5, 1, 5$, and $10 $ au. We set $b_{\rm max}=100\ \rm au$ aimed at the centre of mass of the binary, and the sampling of all the parameters is the same as in the circumstellar case. We restrict to $10^5$ at most for each planetesimal size, while the tightest case of $0.1\ \rm au$ has at most $10^4$ runs, which is the most time consuming due to the need to resolve the binary's orbit (see Table \ref{circumbinarytable} for the full set-up).

Figure \ref{fig:CBfracs} shows the capture fractions for circumbinary discs of different separations, compared to the single case. The gaps in both panels of the plot (missing dots) are from runs where no captures or temporary captures occurred. The left panel shows that the circumbinary separation of $0.1$ au captures more than the single case for all planetesimal sizes. This is due to the total mass enclosed of $2M_\odot$ in the circumbinary case, leading to enhanced gravitational focusing, in particular larger planetesimals. No captures occur beyond $100\ \rm m$ for $5$ and $10$ au binary separations, due to the lack of gas in the inner cavity where most of these captures ought to occur. 

Some temporary captures may still occur for large planetesimals on low enough initial $v_\infty$, but the weak gas drag makes them only temporary, and the binary interactions destabilise them. Temporary captures in this case have the opposite trend to circumstellar ones, increasing with planetesimal radius instead of decreasing due to the weak gas drag. Temporary captures also tend to be less common for tighter binary separations which have smaller cavities and larger gaseous discs.

Figure \ref{fig:CBCDF} shows the final semi-major axis values of planetesimals captured in circumbinary orbits. The truncation radius (now an inner limit) is again plotted. 
We see a clear limit where close to the cavity the smaller planetesimals are significantly deflected and are unable to be captured, and the vast majority are captured in the outer outskirts of the disc. Capture of large planetesimals is rare and occurs mostly near the cavity where the gas surface density is highest.

The circumbinary case clearly demonstrates the need for a dense inner disc when capturing large radii planetesimals. Only tight binaries of separations $\sim 0.1$ au can effectively capture large planetesimals, which is consistent with the higher occurrence of planets around tight binaries. 

\section{Discussion}\label{sec:4}

\subsection{Comparison with observed circumstellar planets}

\begin{figure}
    \centering
    \includegraphics[width=0.99\linewidth]{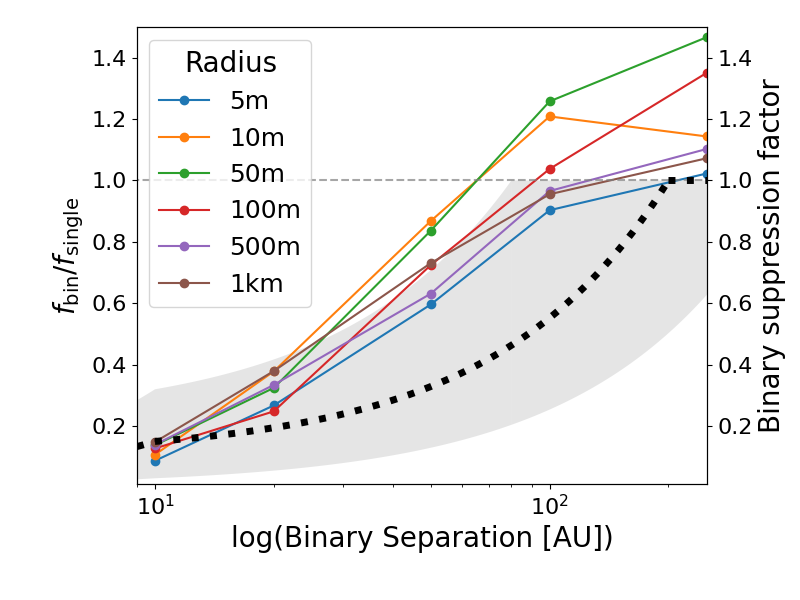}
    \caption{The fraction of binary captures over the fraction of single captures plotted against the binary separation value for the circumstellar case. The horizontal dashed line indicates where binary and single captures are equal. The thick black dotted line and the grey area are the binary suppression factor and its error as calculated by \citet{maxwell2021}.}
    \label{fig:circumstellardivfrac}
\end{figure}

Figure \ref{fig:circumstellardivfrac} shows the fraction of planetesimal capture of different sizes normalised to single case as a function of binary separation. We do not plot the smaller planetesimals below $5\ \rm m$ since they are mostly captured in the geometric regime\footnote{We find the the smallest pebbles of $0.1\ \rm m$ follow the relation of $f_{\rm bin}/f_{\rm sin} \sim (a_{\rm trunc}/b_{\rm max})^2$ where $a_{\rm trunc}$ is the disc truncation radius. This is expected in the purely geometric regime where capture of small pebbles is highly efficient.} and do not affect the subsequent stage of planet formation, which requires larger planetesimals. The thick dotted line (and the grey error margins) overlays an observationally motivated model that provides the suppression factor of circumstellar planets due to a binary companion at a given separation \citep{maxwell2021}. Unsuppressed planet formation is found at larger separations where the suppression factor is $\sim 1$, where the strongest suppression is found around $10\ \rm au$ of about $15\%$.

The general trend of more suppressed planetesimal capture and planet formation for tighter binaries could hint that planetesimal capture is correlated to successful subsequent planet formation. We caution, however, that this trend is only indicative, and many other processes are involved in planet formation. Additionally, our model is limited to two sun-like stars, while \cite{maxwell2021} provides statistics various stellar types and masses. This trend of suppressed planetesimal capture for tighter binaries needs to be tested with future work that explore the stellar masses and disc sizes and structure.

Both reductions in the observed planetary occurrence and planetesimal capture in tighter binaries essentially originate from the same difficulty of planet formation in binary systems. Namely, the associated discs are truncated in size and smaller in mass. If planetesimal capture is an instrumental process for ubiquitous planet formation (individual planets will form on a subset of systems where the conditions for the streaming instability is optimal), it aligns well with the observed population of circumstellar planets.

\subsection{Comparison to observed circumbinary planets}
\begin{figure}
    \centering
    \includegraphics[width=1\linewidth]{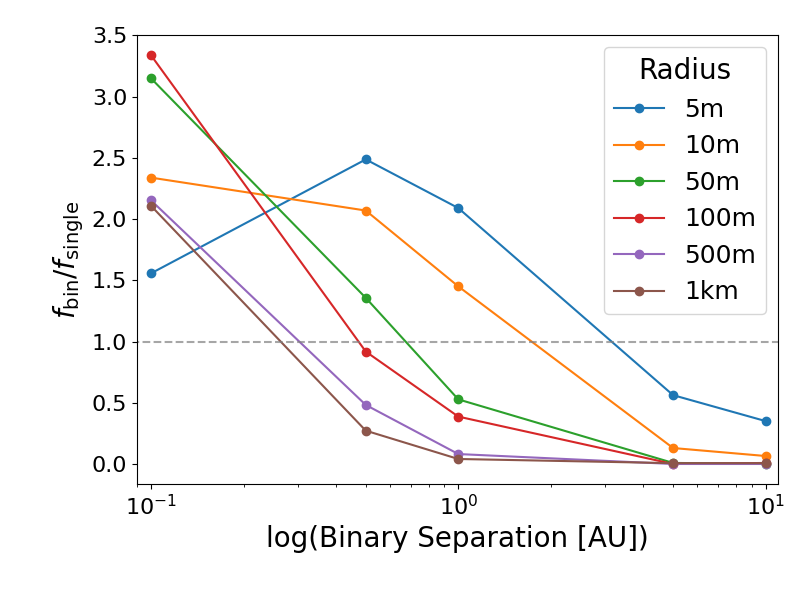}
    \caption{The fraction of binary captures over the fraction of single captures plotted against the binary separation value for the circumbinary case. The horizontal dashed line indicates where the binary captures equal to the single case.}
    \label{fig:circumbinarydivfrac}
\end{figure}

Circumbinary systems show the opposite trend to circumstellar ones (Figure \ref{fig:circumbinarydivfrac}). For tight binaries, capture is more efficient than in the single-star case due to gravitational focusing. As the separation increases, however, inner disc truncation eventually outweighs this benefit, reducing capture efficiency.

Smaller planetesimals retain single-star-like capture efficiencies out to separations beyond  $1$ au, whereas larger planetesimals are affected at separations as small as $0.5$ au. This is because of the preferential capture of larger bodies where the disc density is the highest, which occurs near the inner disc truncation. This trend is consistent with observations showing that most circumbinary planets orbit binaries with separations of $\sim 0.1$ au \cite{thebault2025}, where also a handful of planet exist around binaries with $\mathcal{O}(1\ \rm au)$ separation.

Binarity suppresses capture at all separations, but the effect depends on configuration: wide binaries leave circumstellar capture largely unaffected, while tight binaries similarly preserve circumbinary captures.

\subsection{Total captures per system in a cluster environment}

To further analyse the possibility of capture in realistic binary systems aiding planet formation, we determine how common planetesimal capture is in a cluster environment. The planetesimal encounter rate is given by \citep{gri19} 
\begin{equation}\label{encrate}
    N_\text{enter}\approx n_{\text{ISM}}\sigma_\text{env}\langle v\rangle \tau_\text{env}\text{,}
\end{equation}
where $n_\text{ISM}= n_*N_\text{eject}$ and represents the number density of interstellar planetesimals, while $n_*$ is the number density of stars and $N_{\rm {eject}}$ is the number of planetesimals ejected from each star. $\sigma_\text{env}=\pi b_\text{max}^2$ is the cross section, where $b_\text{max}$ is the maximum impact parameter after which encounters cannot result in captures. $\langle v\rangle=\sqrt{8/\pi}\sigma$ is the mean thermal velocity, with $\sigma$ being the velocity dispersion of the environment. $\tau_\text{env}$ is the typical timescale during which encounters with the disc can occur. Using these relations, Eq. \ref{encrate} can be rewritten as 

\begin{equation}\label{encrate2}
    N_\text{enter}(R)\approx n_*N_\text{eject}(R) b_\text{max}^2\sqrt{8\pi}\sigma \tau_\text{env}\text{,}
\end{equation}

so it is a function of planetesimal radius. The function $N_\text{eject}(R)$ is given by 
\begin{equation}\label{eject}
    N_\text{eject}(R)\sim0.3\left(\frac{4\pi\rho_\text{p}R_\text{p}^3}{3M_\oplus}\right)^{-5/6}\text{,}
\end{equation}
which is based on work done in \cite{gri19}. 
Eq. \ref{encrate2} gives an estimate for the total amount of planetesimals that enter discs within an environment. As we are using a cluster environment for all models, we consider a cluster. With this value for how many planetesimals enter, we can multiply by the capture fraction to convert our capture fractions into number of total captures in an environment. 

This formula, however, is made for the single star case, and therefore must be modified in order to produce results for binaries. 
We take the binary star fraction to be $50\%$, compatible with the binary fraction of FGK-type stars \citep{offner23}. In our simulations, all the stars are solar-like G stars, which falls into the typical planet hosting binary properties. We also must consider how many of those binaries have separations that match those we used in simulations. To do this, we use {\"O}pik's law \citep{Opik} which states that binary separations are log uniformly distributed. {\"O}pik's law as a probability distribution function is given by $dN/da\propto 1/a$.

For simplicity, we consider a range of separations from $2\times 10^{-2}$ au to $2\times 10^3$ au. We further take typical values for an open cluster ($\tau_\text{env}=0.3\text{ Myr}$, $\sigma=6.2\text{ km s}^{-1}$, $n_\star=7.5 \text{ pc}^{-3}$), as well as our simulation values ($b_\text{max}=250$ au for the circumstellar case and $100$ au for the circumbinary case, $\rho_\text{p}=1\text{ g cm}^{-3}$).

Combining the above estimates, Eq. \ref{encrate2} becomes 
\begin{equation}\label{circumbinaryencrate}
     N_{\rm enter}(R) \approx3.64\cdot10^{4}\times\left(\frac{R}{\rm km}\right)^{-5/2} \ln \left(\frac{a_{\rm max}}{a_{\rm min}} \right) 
\end{equation}
for the circumstellar case, where the binary fraction of $50\%$ was assumed. We can consider these circumstellar binaries to be in three different categories: tight ($2-20$ au), medium ($20-200$ au) and wide ($200-2000$ au). The innermost binaries below $2$ au are considered as circumbinary hosts. Choosing a representative case for each category ($10$ au for tight, $50$ au for medium and $250$ au for wide), we obtain the values shown in Table \ref{circumstellartable}. 

\begin{table}
\centering
\begin{tabular}{lllllll}
\hline
Radius (m) & 10 au & 50 au & 250 au & Single Case \\
\hline
10    &$2.06\times10^6$&$1.67\times10^7$&$2.20\times10^7$&$1.93\times10^8$\\
50    &$1.86\times10^4$&$1.13\times10^5$&$1.98\times10^5$&$1.35\times10^6$\\
100   &$2.60\times10^3$&$1.48\times10^4$&$2.76\times10^4$&$2.05\times10^5$\\
500   &$25$&$114$&$199$&$1.81\times10^3$ \\
1000  &$4$&$18$&$26$&$244$\\
\hline
\end{tabular}

\caption{Approximate numbers of planetesimals with different radius values that would have been captured in binary systems with circumstellar discs of different semi-major axis separations, compared with the single case. All values are presented to 3 significant figures.}
\label{circumstellartable}
\end{table}

Table \ref{circumstellartable} shows that the tightest binary of $10$ au can capture $\sim \mathcal{O}(1)$ km planetesimals (and $\sim \mathcal{O}(10)$ for wider binaries). The general scaling of successful capture with size is $\propto N_{\rm eject} \times f_{\rm bin}$. From Eq. \ref{eject}, $N_{\rm eject} \propto R^{-5/2}$, and the capture fraction scaling in the gravitationally focussed regime is similar to the single case, $f_{\rm bin} \propto R^{-2/5}$. Thus, the total scaling law for successful captured planetesimals is $R^{-29/10}$, leading to an estimate of $\sim \mathcal{O}(10^3)$ $100\ \rm m$ captured planetesimals  in cluster environments. We conclude that planetesimal capture in binary systems can still be efficient. If $\sim \rm km$ sized planetesimals are required for successful planet formation, the reduced capture fractions for tight binaries at $10$ au is consistent with the observed circumstellar planet occurrence rates. 

We note that our results are not too sensitive to the cluster properties. Our planetesimal velocity dispersion of $\approx 6.2\ \rm km\ s^{-1}$ is due to the internal mechanisms of planetesimal ejections \citep{Adams05} and it usually exceeds the cluster velocity dispersion, unless the cluster is a massive young stellar cluster of $M_{\rm cl} \gtrsim 10^5 M_\odot$ \citep{spz10}. Such massive clusters are rare, since the initial cluster mass function scales as $dN/dM_{\rm cl} \propto M_{\rm cl}^{-2}$ \citep{lada03, spz10}, hence the majority of planet hosting stars were born in more smaller open clusters and associations. The encounter rate with neighbouring stars is much longer than the typical disc lifetime \citep{spz10}, so it shouldn't affect our estimates. Varying the binary mass ratio may result in predominant capture onto the primary disc. In this case, the capture rates will resemble the single case onto the more massive star. Detailed examination of varying mass and mass ratios is beyond the scope of this paper.

\subsection{Planetesimal capture in the broader context of planet formation}
The total captured mass is dominated by the most massive object captured, which is $\sim (4\pi/3)\rho_p R^3 \sim 10^{-10} M_\oplus$ for a $\sim 5\ \rm km$ planetesimal. Although the total captured mass is low, it already arrives in an assembled planetesimal form. These km-sized planetesimals are resistant against radial drift and are long lived. Although pebble accretion and/or runaway growth is generally efficient for much larger embryos ($\sim 100\ \rm km$), the km-sized planetesimals can still grow by direct accretion of $<$ cm particles, known as the 'snowball' phase \citep{snowball}. Moreover, the snowball phase is more efficient if the planetesimals are are on isolated orbit and avoid destructive mutual collisions. If a fraction $\epsilon_p$ of the dust is in the form of planetesimals at $t_0 \sim 10^5$ yr (typical for cluster environment), this corresponds to a long planetesimal formation timescale of $t_f=10^5/\epsilon_p \rm \ yr$, which is defined where most of the dust is in planetesimal mass (see Eq. 2 of \citealp{snowball}). For out choice of parameters, for a total $10M_\odot$ of solid mass, $\epsilon_p=10^{-11}$ and $t_f=10^{16}$ yr. This corresponds to the upper limit of figures 2-4 of \cite{snowball} that show that the initial km-sized planetesimals can form $> 100$ km embryos within a Myr for weak and moderate turbulence (and $\sim 40$ km planetesimals within a few Myr for strong turbulence).

In this paper we focus only on the capture of interstellar planetesimals in binary systems. We remain agnostic to how and where the first planetesimals forms (which can be a rare event in one of the more massive and/or warped protoplanetary disc in the cluster, or in a different older cluster when the planetesimals had time to scatter and escape the cluster and become interstellar planetesimals). We also remain agnostic to the other formation and growth models. While runaway growth and pebble accretion can be valuable for growing the $\sim 100$ km embryos to planetary cores, the planetesimal capture provides the first seeds for these planetesimals. If the streaming instability is not omnipresent in protoplanetary discs, additional mechanisms such as planetesimal capture and growth via the snowball phase may provide the necessary planetesimal population for subsequent growth.

\subsection{Caveats and limitations}

 We ran the circumstellar case for around $10^5-10^6$ simulations per point to obtain the capture fractions in Figure \ref{fig:physicalcapturefraction}, with more runs being done for the planetesimals that were captured less to allow better statistics. However, the circumbinary model posed problems due to the computational cost of simulating close binaries. The statistics for this model could be improved by increasing the number of simulations run.

In order to simplify and isolate the role of binarity, we focussed on equal-mass circular binaries of $1M_\odot$ stars. Although the initial mass-function (IMF) has a relatively step slope, the planet-hosting mass function is heavily centred near $1\ M_\odot$ in FGK-type stars due to targeted radial-velocity searches \cite[e.g.,][]{mayer14, zink23}. The exoplanet population around M-dwarfs is expected to be incomplete due to observational biases. Our simulations do not account for the spread of binary masses and eccentricities, both factors that would effect the capture rate of planetesimals. The binary fraction of M-dwarfs is also reduced to $\sim 20\%$ \citep{offner23} which could affect our rate estimate by a factor of a few. Further investigation could be done on the effects of varying the mass ratio and eccentricity, as both contribute to disc truncation.

Our disc models are also simplified power law density profiles and are aligned with the binary orbital plane, and the diversity of protoplanetary disc structure and orientations (as up to 75\% of protoplanetary discs may be warped \citealp{Villenave2024}) should be explored in future work. The lack of disc diversity and the consistent disc mass and density profile used throughout does not reflect the reality of the protoplanetary disc population, and could limit the results. Discs with larger masses would exert larger drag force, and vice versa for smaller masses, all considerations that are absent from our results for simplicity.

Finally, due to the N-body nature of our investigation, the disc is a static external force in the background and does not have feedback or evolve. Planetesimals on eccentric orbits can lead to chondrule formation, crystallization of silicate dust, and evaporation of icy planetesimals \citep{nagasawa19}, thereby altering the disc composition. While individual heating events are likely too localized and short-lived to be directly observable, their cumulative effects may produce detectable chemical or thermal asymmetries in protoplanetary discs. Additional features in protoplanetary discs such as kinks, warps, arc, gaps etc. \citep[e.g.,][]{Nealon20, Rowther25} need to be coupled to hydrodynamic modelling and dust growth in order to further quantify the robustness of planetesimal formation and capture.

\section{Summary}\label{sec:5}

We explored the possibility of planetesimal capture in binary star systems, considering both the circumstellar and circumbinary cases. Our main conclusions are as follows:

i) In the circumstellar case, the capture probability of large ($>50$ m) planetesimals is reduced for binary separations below $\sim 50$ au, with the strongest suppression occurring at $10$ au, where the capture fraction decreases by approximately an order of magnitude (Fig. \ref{fig:CDFphysical}). The relative reduction in capture efficiency compared to the single-star case broadly matches the observed reduction in planet occurrence in binary systems (Fig. \ref{fig:circumstellardivfrac}).

ii) The circumbinary case is substantially more detrimental for capture. While the tightest binaries ($\sim 0.1$ au) exhibit capture rates comparable to the single-star case, binaries with separations $\gtrsim 1$ au capture only $\sim 1\%$ as many large planetesimals. The widest binaries are unable to capture planetesimals larger than $\sim 50$ m (Figs. \ref{fig:CBfracs} and \ref{fig:circumbinarydivfrac}). This trend is also consistent with the observed properties of circumbinary planets, which predominantly orbit binaries with separations near $0.1$ au, with only a small number extending to $\sim 1$ au.

iii) Similar to the single-star case, larger planetesimals are preferentially captured in the inner disc regions due to the enhanced surface density (Figs. \ref{fig:CDFphysical} and \ref{fig:CBCDF}). In the circumstellar case, the binary introduces a dynamical instability interior to $a_c \approx 0.27a_b$ for equal-mass circular binaries \citep{holman1999}. As a result, many small planetesimals cannot be captured in the region between the instability boundary and the disc truncation radius, $a_{\rm trunc}\approx 0.47a_b$. Similar behaviour is observed in the circumbinary case.

iv) Tight binaries ($\sim 10\ \rm au$) in cluster environments can capture $\mathcal{O}(1)$ km-sized planetesimals, while wider binaries can capture up to $\mathcal{O}(10)$ such objects. If planet formation requires the capture of $\gtrsim 1\ \rm km$ planetesimals, our results remain broadly consistent with the observational constraints on planet occurrence in binaries.

Planet formation in binaries remains poorly understood, and identifying mechanisms that help overcome its additional dynamical challenges is therefore important. Our results show that planetesimal capture is a potentially viable component of planet formation in binary systems. Although the total captured mass is much smaller than the initial dust mass, the material is already assembled into planetesimals, which may further facilitate processes such as snowball and runaway growth, streaming instability and pebble accretion. While the overall picture of planet formation in binaries is far from complete, planetesimal capture appears to be a viable contributing mechanism that may help explain the observed population of planets in binary systems.

\section*{Acknowledgements}
We thank Rebecca Nealon for giving feedback on the content and structure of this paper. We thank the anonymous referee for helpful feedback that has improved this paper. We thank Gabriel McMahon-Killingsworth and Alexander Heger for stimulating discussions. EG acknowledges support from
the ARC Discovery Early Career Research Award (DECRA) DE260101802. We acknowledge the CPU time on OzSTAR and Ngarrgu Tindebeek funded by the Victorian and Australian
governments and Swinburne University.

\section*{Data Availability}
The scripts used for running Rebound are available at \href{https://github.com/AliaPollard/Honours-Project}{https://github.com/AliaPollard}.
The data underlying this article will be provided by the corresponding author upon reasonable request.

\bibliographystyle{mnras}

\begin{table*}
\begin{tabular}{ccccccc}
\hline 
\multicolumn{1}{|c|}{Size {[}m{]}} & \multicolumn{1}{c|}{$10$ au} & \multicolumn{1}{c|}{$20$ au} & \multicolumn{1}{c|}{$50$ au} & \multicolumn{1}{c|}{$100$ au} & \multicolumn{1}{c|}{$250$ au} & \multicolumn{1}{c|}{Single}\tabularnewline
\hline 
\multicolumn{7}{|c|}{Comparative model ($b_{{\rm max}}=100\ {\rm au})$}\tabularnewline
\hline 
$0.1$ & $124/10^{4}$ & $212/10^{4}$ & $619/10^{4}$ & $1460/10^{4}$ & $4997/9.1\cdot10^{3}$ & $9114/10^{4}$\tabularnewline
$0.5$ & $94/10^{4}$ & $138/10^{4}$ & $405/10^{4}$ & $1349/10^{4}$ & $1535/3.1\cdot10^{3}$ & $8603/10^{4}$\tabularnewline
$1$ & $66/2\cdot10^4$ & $89/2\cdot10^4$ & $154/2\cdot10^4$ & $701/2\cdot10^4$ & $3136/1.56\cdot10^4$ & $38306/10^{5}$\tabularnewline
$5$ & $6/10^{5}$ & $12/10^{5}$ & $126/10^{5}$ & $165/10^{5}$ & $1114/10^{5}$ & $2454/10^{5}$\tabularnewline
$10$ & $7/10^{5}$ & $23/10^{5}$ & $87/10^{5}$ & $117/10^{5}$ & $629/10^{5}$ & $1227/10^{5}$\tabularnewline
$50$ & $3/10^{5}$ & $19/10^{5}$ & $85/10^{5}$ & $56/10^{5}$ & $264/10^{5}$ & $468/10^{5}$\tabularnewline
$100$ & $2/10^{5}$ & $21/10^{5}$ & $84/10^{5}$ & $32/10^{5}$ & $189/10^{5}$ & $354/10^{5}$\tabularnewline
$500$ & $4/10^{5}$ & $17/10^{5}$ & $70/10^{5}$ & $19/10^{5}$ & $85/10^{5}$ & $197/10^{5}$\tabularnewline
$1000$ & $4/10^{5}$ & $23/10^{5}$ & $68/10^{5}$ & $21/10^{5}$ & $69/10^{5}$ & $147/10^{5}$\tabularnewline
\hline 
\multicolumn{7}{|c|}{Circumstellar model ($b_{{\rm max}}=250\ {\rm au})$}\tabularnewline
\hline 
$0.1$ & $230/10^{5}$ & $607/10^{5}$ & $1716/10^{5}$ & $430/10^{4}$ & $1138/9.8\cdot10^3$ & $5718/10^{4}$\tabularnewline
$0.5$ & $216/10^{5}$ & $533/10^{5}$ & $1886/10^{5}$ & $446/10^{4}$ & $988/8.4\cdot10^{3}$ & $3649/10^{4}$\tabularnewline
$1$ & $159/10^{5}$ & $475/10^{5}$ & $1453/10^{5}$ & $314/10^{4}$ & $594/9.9\cdot10^3$ & $1279/10^{4}$\tabularnewline
$5$ & $283/6.16\cdot10^5$ & $1409/10^{6}$ & $3111/9.9\cdot10^5$ & $476/10^{5}$ & $539/10^{5}$ & $527/10^{5}$\tabularnewline
$10$ & $246/10^{6}$ & $870/10^{6}$ & $1997/10^{6}$ & $278/10^{5}$ & $263/10^{5}$ & $230/10^{5}$\tabularnewline
$50$ & $124/10^{6}$ & $292/10^{6}$ & $753/10^{6}$ & $1132/10^{6}$ & $132/10^{5}$ & $90/10^{5}$\tabularnewline
$100$ & $47/4.8\cdot10^5$ & $191/10^{6}$ & $558/10^{6}$ & $799/10^{6}$ & $104/10^{5}$ & $77/10^{5}$\tabularnewline
$500$ & $53/10^{6}$ & $127/10^{6}$ & $240/10^{6}$ & $367/10^{6}$ & $419/10^{6}$ & $38/10^{5}$\tabularnewline
$1000$ & $43/10^{6}$ & $110/10^{6}$ & $212/10^{6}$ & $277/10^{6}$ & $311/10^{6}$ & $29/10^{5}$\tabularnewline
\end{tabular}

\caption{Numbers of captures and simulations for the comparative and circumstellar models. The single case listed is specific to the $b_{\rm{max}}$ used for each model.}
\end{table*}

\begin{table*}
\begin{tabular}{ccccccc}
\hline 
\multicolumn{1}{|c|}{Size {[}m{]}} & \multicolumn{1}{c|}{$0.1$ au} & \multicolumn{1}{c|}{$0.5$ au} & \multicolumn{1}{c|}{$1$ au} & \multicolumn{1}{c|}{$5$ au} & \multicolumn{1}{c|}{$10$ au} & \multicolumn{1}{c|}{single}\tabularnewline
\hline 
\multicolumn{7}{|c|}{Circumbinary model ($b_{{\rm max}}=100\ {\rm au})$}\tabularnewline
\hline 
$0.1$ & $9342/9.9\cdot10^3$ & $18946/2\cdot10^{4}$ & $9464/10^{4}$ & $9349/10^{4}$ & $9057/10^{4}$ & $9114/10^{4}$\tabularnewline
$0.5$ & $9026/9.7\cdot10^3$ & $18781/2\cdot10^{4}$ & $9390/10^{4}$ & $9274/10^{4}$ & $8956/10^{4}$ & $8603/10^{4}$\tabularnewline
$1$ & $937/1.7\cdot 10^3$ & $12069/2\cdot10^{4}$ & $6156/10^{4}$ & $5609/10^{4}$ & $5093/10^{4}$ & $38306/10^{5}$\tabularnewline
$5$ & $363/9.5\cdot10^3$ & $1221/2\cdot10^{4}$ & $513/10^{4}$ & $138/10^{4}$ & $86/10^{4}$ & $2454/10^{5}$\tabularnewline
$10$ & $287/10^{4}$ & $508/2\cdot10^{4}$ & $178/10^{4}$ & $16/10^{4}$ & $8/10^{4}$ & $1227/10^{5}$\tabularnewline
$50$ & $143/9.7\cdot10^3$ & $127/2\cdot10^{4}$ & $247/10^{5}$ & $4/10^{5}$ & $1/10^{5}$ & $468/10^{5}$\tabularnewline
$100$ & $117/9.9\cdot10^3$ & $65/2\cdot10^{4}$ & $137/10^{5}$ & $0/10^{5}$ & $3/10^{5}$ & $354/10^{5}$\tabularnewline
$500$ & $42/9.9\cdot10^3$ & $19/2\cdot10^{4}$ & $16/10^{5}$ & $0/10^{5}$ & $0/10^{5}$ & $197/10^{5}$\tabularnewline
$1000$ & $31/10^{4}$ & $8/2\cdot10^{4}$ & $6/10^{5}$ & $1/10^{5}$ & $1/10^{5}$ & $147/10^{5}$\tabularnewline
\end{tabular}

\caption{Numbers of captures and simulations for the circumbinary model.}
\label{circumbinarytable}
\end{table*}
\appendix

\section{Number of simulations and captures for each model}\label{appen:a}

The following tables lists the number of captures and the number of total successful simulations for the single, comparative, circumstellar and circumbinary models. The choice for $b_{\rm max}$ is listed before each set-up. Some simulations did not finish for the time limit given on the HPC and are not listed.
\end{document}